# Non-Hermitian physics and engineering in silicon photonics

Changqing Wang, Zhoutian Fu, and Lan Yang*

*Department of Electrical and Systems Engineering, Washington University, St. Louis, MO 63130, USA*
*\*yang@seas.wustl.edu*

Silicon photonics has been studied as an integratable optical platform where numerous applicable devices and systems are created based on modern physics and state-of-the-art nanotechnologies. The implementation of quantum mechanics has been the driving force of the most intriguing design of photonic structures, since the optical systems are found of great capability and potential in realizing the analogues of quantum concepts and phenomena. Non-Hermitian physics, which breaks the conventional scope of quantum mechanics based on Hermitian Hamiltonian, has been widely explored in the platform of silicon photonics, with promising design of optical refractive index, modal coupling and gain-loss distribution. As we will discuss in this chapter, the unconventional properties of exceptional points and parity-time symmetry realized in silicon photonics have created new opportunities for ultrasensitive sensors, laser engineering, control of light propagation, topological mode conversion, etc. The marriage between the quantum non-Hermiticity and classical silicon platforms not only spurs numerous studies on the fundamental physics, but also enriches the potential functionalities of the integrated photonic systems.

## 1. Introduction

Non-Hermitian physics describes open quantum systems that have interactions with environment in the form of matter or energy exchange. In quantum mechanics, the Hamiltonian ($H$) describing a closed quantum system is a Hermitian operator satisfying $H^\dagger = H$ [1]. Its eigenvalues are real, and the corresponding eigenstates are orthogonal, which provide a complete basis to describe any state of the system in the Hilbert space. However, such argument conveys no information of the spectral behavior and state description when the system interacts – often unavoidably – with the environment. The most common form of interaction is the energy dissipation from the system to the ambient environment, such as decoherence in atomic states, or optical absorption/radiation/scattering loss in confined structures. Even though the density matrix description and the master equation approach [2–4] offer standard tools for describing behavior of open systems in different situations, in many cases, concise descriptions directly offered by Hamiltonians are more favorable.

Towards the end of the 20th century, the traditional scenario of quantum theory was expanded by introducing non-Hermitian Hamiltonian, which satisfies $H^\dagger \neq H$, to describing open systems [5–7]. The eigenvalues are in general complex, and the associated eigenstates can be non-orthogonal. In particular, non-Hermitian degeneracies happen at an exceptional point (EP) where two or more eigenvalues and corresponding eigenstates coalesce simultaneously. More interest was aroused to a particular family of open systems preserving parity-time (*PT*) symmetry [8–14], whose Hamiltonian commutes with the joint operation of parity operator (*P*) and time operator (*T*), *i.e.*, $[H, PT] = 0$. *PT*-symmetric systems are found to exhibit real eigenspectra out of complex potential under certain range of parameters, while the eigenstates keep non-orthogonal to each other. Theses extraordinary properties of systems with dissipation (or amplification) arouse a series of fundamental and engineering studies in various quantum systems.

About a decade later, non-Hermitian physics sent a new wave of ripples in classical systems [13], including electronics [15,16], metamaterial [17], acoustics [18], and especially optics [19–24]. Helmholtz equations under paraxial approximation can be viewed in analogy to the Schrödinger equation, with the quantum potential replaced by optical index profiles [25]. The real and imaginary parts of the potential can be realized by optical refractive index and the optical gain/loss, respectively, which are quite feasible with the modern nanophotonic technique. As a result, a series of studies have unveiled unconventional properties of optical systems operating around EPs, including spectral singularity, chiral modes, unidirectional reflection and so on, which bring new opportunities to sensing, mode engineering and optical signal processing. Interest in *PT*-symmetry also boomed due to its potential applications in engineering on-chip lasers and light transport, such as single mode lasing and nonreciprocal light propagation.

Silicon photonics has played an indispensable role in the progress of non-Hermitian physics in optical systems. They offer unique advantageous features for optical design, including integratability, low cost, scalability, room-temperature operation, flexible control of optical parameters (field intensity, phase, polarization, refractive indices, mode coupling strengths and optical damping rates), compatibility with other degrees of freedom, *etc.* [26]. Especially, many ideas were implemented into the optical structures such as waveguides and resonators, as well as photonic crystals. In such structures, optical resonant frequencies can be controlled via the thermo-optic effect [27,28], while the coupling between different units can be introduced and adjusted by spatial positioning [29], backscattering engineering [29,30], or geometric deformation [31–34]. The optical gain can be introduced by doping rare-earth ions in silica material [35], while loss can be engineered by external absorbers [36] or the coupling loss induced by incoming/outcoming channels [37]. In both waveguides and resonators, the spatial deposition of absorbing material such as Germanium (Ge) and Chromium (Cr) onto silicon-based structures enable local engineering of refractive indices. In addition, by coating silica optical devices with other transparent optical material such as polymer [38–42] or silk fibroin [43,44], one can bring modification to optical, thermal and mechanical properties and render novel nonlinear behavior. As a result, the silicon photonics platforms provide a fertile ground to the studies of non-Hermitian physics. In return, non-Hermitian physics unleashes the full potential of silicon photonics for novel functionalities and applications spanning from sensing, lasing and optical information processing to the control of nonlinear optical behavior.

In this chapter, we review the study of non-Hermitian physics and the associated engineering progress in silicon photonics. The chapter is organized as follows. In section 2, we introduce the theory on how non-Hermitian physics can be realized in optical systems in an analogous manner to quantum scenarios. In section 3, we target on the eigenspectrum of non-Hermitian system, where we discuss the spectral singularity associated with the exceptional points, and how it leads to ultrasensitive sensors. In section 4, we review the studies about the mode interaction and eigenstate features in non-Hermitian optical systems. In particular, the supermodes in *PT*-symmetric systems and the chiral eigenmodes at EPs enable a series of unconventional lasing behavior. In section 5, we turn to the optical scattering properties enabled by EPs and PT symmetry, i.e., the system response when probed by optical waves. Such properties exert significant influence on various light propagation features including directionality, nonreciprocity and group delay. The topological properties associated with the EP eigenspectrum are discussed in section 6, which paves new way in the control and switch of optical modes by topological operations. Finally, we conclude the chapter by offering an outlook in section 7.

## 2. Non-Hermitian physics: from quantum mechanics to optics

### 2.1  Non-Hermitian physics in quantum mechanics

Non-Hermitian quantum mechanics is an alternative formalism to standard Hermitian quantum mechanics and presents a unique approach to describe open quantum systems. Non-Hermitian theory is introduced to quantum mechanics mainly for two reasons [5]. First, in quantum mechanics, many problems can be treated more easily, either analytically or numerically, by non-Hermitian quantum mechanics than using the standard quantum theory. Second, in many problems, such as classical mechanical statistics and propagation of electromagnetic waves, we can describe these cases of interest with a Hamiltonian that is in analogy to the quantum description. The analogous Hamiltonian, however, may not respect Hermiticity, and thereby in many cases, the analogous quantum problem can only be described by non-Hermitian theory and have no counterpart in standard Hermitian quantum mechanics. In this chapter, we will mainly focus on the second motivation with phenomena found in silicon photonics being the main topic of interest. More interesting phenomena unveiled by non-Hermitian physics can be found in related books and reviews [5,13].

One of the most important ways of introducing non-Hermitian formalism to quantum systems is to consider complex potentials in Schrödinger equations. It is known that in standard quantum mechanics the potential can only be a real function, conserving the total particle numbers or stored energy, while the non-conservative open features can only be depicted by more complicated formalism using density matrix approach. Non-Hermitian quantum theory makes sense of the complex potential that breaks Hermiticity. To be more specific, the Schrödinger equation describing the evolution of a quantum wavefunction $\Psi(\boldsymbol{r},t)$ in a time-invariant complex potential is given by

$$i\hbar \frac{\partial}{\partial t}\Psi(\boldsymbol{r},t) = \left[-\frac{\hbar^2}{2m}\nabla^2 + V(\boldsymbol{r})\right]\Psi(\boldsymbol{r},t), \qquad (1)$$

where the complex potential $V(\boldsymbol{r})$ implies non-conserved probability $\int |\Psi(\boldsymbol{r},t)|^2 d\boldsymbol{r}$.

Non-Hermitian quantum theory not only offers an alternative approach for open quantum problems, but also introduces unconventional physical concepts and phenomena, such as self-orthogonal eigenstates, parity-time symmetry, etc. One can find considerable disparity between Hermitian and non-Hermitian systems. Closed systems with Hermitian Hamiltonians always have real eigenvalues and orthogonal eigenstates under which the Hamiltonians are represented by diagonalized matrices. Their eigenstates $|\psi_i\rangle$ are orthonormal under the inner product, *i.e.*, $\langle\psi_i|\psi_j\rangle = \delta_{i,j}$. Non-Hermitian systems, however, can possess complex eigenvalues and non-orthogonal eigenstates. Let us consider a non-Hermitian operator $A$ with eigenvalues $\lambda_i$ and the corresponding eigenvectors $|\phi_i\rangle$, *i.e.*, $A|\phi_i\rangle = \lambda_i|\phi_i\rangle$. In order to find an orthonormal eigenbasis, we define bi-orthogonal product (c-product) in replacement of the inner product, such that $(\phi_i|\phi_j) = \delta_{i,j}$ [5].

A peculiar situation occurs at non-Hermitian degeneracies where two or more non-orthogonal eigenvectors coalesce, which is also named as branch point, or EP. The bi-orthogonality yields $(\phi_{EP}|\phi_{EP}) = 0$, i.e., the eigenstate at an EP is self-orthogonal. Such self-orthogonality is responsible for a great many singular phenomena, including divergence of expectation values of non-Hermitian operators, creation of probabilities, Berry phase accumulation by adiabatic encircling EPs, and so on [5].

On the other hand, the complex potential can generate purely real eigenvalues, if preserving symmetry under a joint operation of parity ($P$) and time-reversal ($T$) operators, known as parity-time ($PT$) symmetry [8]. The condition for a parity-time symmetric potential is given as

$$PTV(\mathbf{r}) = V(\mathbf{r}), \qquad (2)$$

which yields a requirement for the complex potential

$$V(\mathbf{r}) = V^*(-\mathbf{r}). \qquad (3)$$

So far, we have provided the basic concepts of non-Hermitian quantum mechanics, including the definitions of EPs and $PT$ symmetry, in the context of open quantum systems. Below we show how these quantum concepts and the associated phenomena can be readily mimicked and applied in classical optical systems. We classify the optical realization of non-Hermitian physics including $PT$ symmetry and EPs into three categories: wave propagation under paraxial approximation in a transverse complex potential, wave scattering in a longitudinal complex potential and discrete optics with a matrix form of Hamiltonian.

## 2.2  Paraxial propagation of electromagnetic field in a transverse complex potential

The classical realization of non-Hermitian physics can be first seen by investigating the propagation of electromagnetic waves in an optical medium with planar structures [25,45,46]. We start our analysis from a one-dimensional case shown in Fig. 1a, where the field propagation occurs in a 1D potential along $z$ axis and uniformity is assumed for the potential in the $y$ direction. Without loss of generality, we consider a TE wave with electric field $\mathbf{E}(x,z)e^{-i\omega t}$, for which the Helmholtz equation is given by

$$[\nabla^2 + k^2 \epsilon(x,z)]\mathbf{E}(x,z) = 0. \qquad (4)$$

If the medium is uniform along $z$ direction, we can rewrite the Helmholtz equation under the paraxial approximation $\mathbf{E}(x,z) \sim E(x)e^{ikz}\hat{e}$, where $\hat{e}$ is the direction of the electric field,

$$ik\frac{\partial}{\partial z}E(x) = -\frac{\partial^2}{\partial x^2}E(x) - \frac{\omega^2}{c^2}\epsilon(x)E(x), \qquad (5)$$

which takes a similar form to that of the Schrödinger equation. The time evolution is replaced by propagation along the $z$ direction. The potential $V(x)$ in Schrödinger equations is represented by a term consisting of the complex permittivity $\epsilon(x)$, or equivalently the squared refractive index $n(x)^2$. Thereby, the propagation of electromagnetic field in paraxial approximation is in natural analogy to the quantum evolution of wavefunction within a 1D complex potential. It is now obvious that the concept of PT symmetry can be introduced to the optical context with the complex potential satisfying

$$\epsilon(x) = \epsilon^*(-x), \tag{6}$$

which can be realized in linear media by designing the refractive index distribution

$$n(x) = n^*(-x), \tag{7}$$

indicating that the complex potential has symmetric real parts and anti-symmetric imaginary parts (Fig. 1b), *i.e.*,

$$\begin{aligned} n_R(x) &= n_R(-x), \\ n_I(x) &= -n_I(-x). \end{aligned} \tag{8}$$

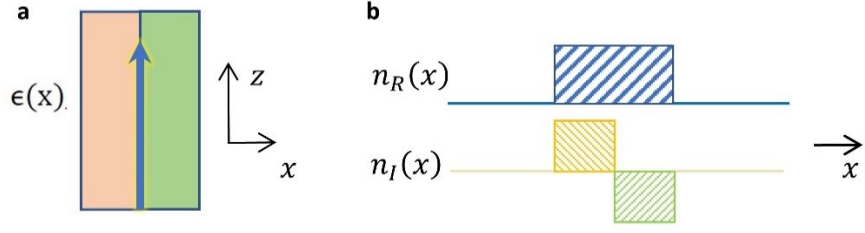

**Fig. 1** Realization of parity-time (**PT**) symmetry in a transverse optical potential. **a** Wave propagation in a planar waveguide which has a distribution of electric permittivity $\epsilon(x)$ in the transverse ($x$) direction. **b** An example distribution of the real and imaginary part of the refractive index in the $x$ direction to mimic a **PT**-symmetric potential.

Similarly, in the two-dimensional potential, the condition for $PT$-symmetry is modified to

$$n(x, y) = n^*(-x, -y). \tag{9}$$

Eqs. (7)-(9) provide the basic guidance for engineering $PT$-symmetry in optics, with the real and imaginary parts of the refractive index governing the oscillating behavior and amplification/dissipation features of the electromagnetic wave. However, as we can see, the above discussion focuses on the situation that the optical potential remains uniform along the direction of wave propagation, which intends to mimic a time-invariant quantum complex potential. A natural question to ask is whether and how a $PT$-symmetric system can be found if the wave encounters variance of refractive index during propagation.

### 2.3 Wave scattering in a longitudinal complex potential

In this section, we turn to the discussion of the physical realization of wave scattering in a longitudinal complex optical potential [47]. For a planar waveguide shown in Fig. 2, the optical potential is modulated along the $z$ axis which is the direction of propagation in contrast to the situation in the last section where the refractive index is only modified transversely. We assume that the linear uniform optical medium has a resonance frequency $\omega_0$, a plasma frequency $\omega_p$ and damping constant $\delta$. The relative permittivity is then given by

$$\epsilon(z, \omega) = 1 - \xi(z) \frac{\omega_p^2}{\omega^2 - \omega_0^2 + 2i\delta\omega}, \tag{10}$$

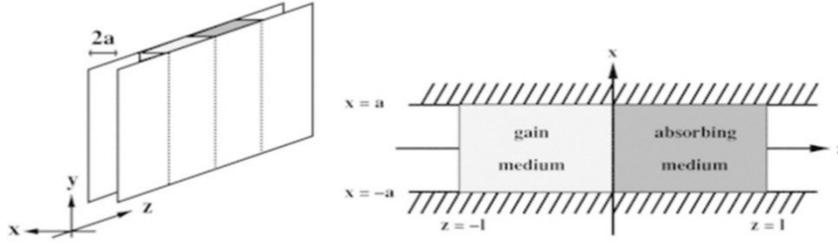

**Fig. 2** Realization of **PT**-symmetry in an optical scattering potential. Schematic diagram for the planar slab waveguide in the three-dimension view and the top view which are shown in the left and right panels, respectively. Figures from Ref. [47].

where $\xi(z)$ regulates the gain/loss distribution along the $z$ direction. For example, the gain medium has $\xi(z) = -1$ and the absorbing medium has $\xi(z) = 1$. In the vacuum, $\xi(z) = 0$.

Considering a single-frequency TM mode with the electric field $E_y(x, y, z, \omega) = -i\omega \hat{B}_x(x)\phi(z, \omega)$ and the magnetic field $B_x(x, y, z, \omega) = \hat{B}_x(x)\frac{\partial \phi}{\partial x}(z, \omega)$, $B_z(x, y, z, \omega) = -\frac{\partial \hat{B}_x}{\partial x}(x)\phi(z, \omega)$, we can derive the following relations based on the Maxwell equations

$$\left(c^2 \frac{\partial^2}{\partial x^2} + \omega_c^2\right)\hat{B}_x(x) = 0, \tag{11}$$

$$\left[\frac{\omega^2 \epsilon(z, \omega) - \omega_c^2}{c^2} + \frac{\partial^2}{\partial z^2}\right]\phi(z, \omega) = 0, \tag{12}$$

with $\omega_c$ being the cut-off frequency.

Now we consider the frequency $\omega$ to be slightly larger than the cut-off frequency $\omega_c$, i.e., $\omega = \omega_c + \Delta\omega$, as well as the assumption $\frac{\omega_p^2}{\delta} \ll \delta, \omega_c, \Delta\omega \ll \delta, \omega_c$. One can find that

$$\frac{\partial^2}{\partial x^2}\phi(z, \omega_c + \Delta\omega) = -\left(i\xi(z)\frac{\omega_c \omega_p^2}{2c^2 \delta} + 2\Delta\omega \frac{\omega_c}{c^2}\right)\phi(z, \omega_c + \Delta\omega). \tag{13}$$

The approximation made so far has reduced all higher order terms of $\Delta\omega/\omega_c$ so that the time evolution term is linearized to a simple form. Reconverting Eq. (13) to the time domain with the definition $\psi(z, t) = \int d\Delta\omega \phi(z, \omega_c + \Delta\omega)e^{-i\Delta\omega t}$, we get

$$i\frac{\partial \psi}{\partial x}(z, t) = -\frac{c^2}{2\omega_c}\frac{\partial^2}{\partial z^2}\psi(z, t) - i\xi(z)\frac{\omega_p^2}{4\delta}\psi(z, t). \tag{14}$$

It is not hard to find that Eq. (14) takes the form of a Schrödinger equation with the definition of the effective mass of photon $m = \frac{\hbar \omega_c}{c^2}$ and the effective potential $V(z) = -i\xi(z)\frac{\omega_p^2 \hbar}{4\delta}$. Therefore, non-Hermitian phenomena can be generated by judiciously designing the optical structures. For example, the condition of *PT* symmetric scattering potential for the mode with near-cutoff frequency is $V(z) = V^*(-z)$. This can be achieved, for instance, by engineering the gain loss distribution in the waveguide to realize a simple form of $\xi(z)$ as

$$\xi(z) = \begin{cases} -1 & \text{gain medium} & -l < z < 0 \\ 1 & \text{absorbing medium} & 0 < z < l \\ 0 & \text{vaccum} & |z| > l \end{cases}. \qquad (15)$$

## 2.4 Non-Hermitian optical waveguides and resonators

Being the most indispensable elements in photonic devices and systems, guided wave optical structures such as waveguides and resonators are common platforms for the realization of non-Hermitian behavior and functionalities. Moreover, the above-mentioned wave propagation in paraxial approximation can be readily applied to optical transport in guided wave structures. Therefore, in this section, we take a further step to introduce the coupled mode theory describing the non-Hermitian guided wave optical structures.

Without loss of generality, let us consider a 1D coupled waveguide system consisting of two waveguides with a difference in their uniformly distributed loss (or gain). By finding the extremum of Lagrangian with respect to the mode amplitude $a_1(z)$ and $a_2(z)$, one can derive the coupled-mode equations [48–50]

$$i\frac{da_1}{dz} = -i\gamma_1 a_1 + \kappa_{12} a_2, \qquad (16)$$

$$i\frac{da_2}{dz} = -i\gamma_2 a_2 + \kappa_{21} a_1, \qquad (17)$$

with $\gamma_1$ and $\gamma_2$ representing the loss rates for the first and second waveguide modes, respectively, and $\kappa$ being the coupling strength between the two modes. It follows that by writing $i\frac{da}{dz} = Ha$, where $a = \begin{pmatrix} a_1 \\ a_2 \end{pmatrix}$, one can derive a Hamiltonian matrix out of the coupled-mode equations [13]

$$H = \begin{pmatrix} -i\gamma_1 & \kappa_{12} \\ \kappa_{21} & -i\gamma_2 \end{pmatrix}. \qquad (18)$$

Note that $H \neq H^\dagger$ implies the non-conservation of energy induced by, for instance, the existence of loss, even though we usually have $\kappa_{12} = \kappa_{21}^*$ for energy-conserved coupling. If one waveguide is engineered to have loss and the other has an equal amount of gain, i.e., $\gamma_1 = -\gamma_2 > 0$, then the system respects $PT$-symmetry.

On the other hand, optical resonators are another set of ideal candidates for implementing non-Hermitian potential. As a counterpart to the spatial propagation dynamics in coupled waveguides, the coupled resonators have dynamic temporal evolution described by temporal-coupled-mode theory [51]

$$\begin{aligned} i\frac{da_1}{dt} &= \left(\omega_1 - i\frac{\gamma_1}{2}\right) a_1 + \kappa_{12} a_2, \\ i\frac{da_2}{dt} &= \left(\omega_2 - i\frac{\gamma_2}{2}\right) a_2 + \kappa_{21} a_1, \end{aligned} \qquad (19)$$

where $\omega_{1,2}$ are resonant frequencies of the mode 1 (with field amplitude $a_1$) and mode 2 (with field amplitude $a_2$), $\gamma_{1,2}$ are intrinsic loss rates of the resonators, and $\kappa_{12}$ ($\kappa_{21}$) is the coupling strength from mode 2 to mode 1 (mode 1 to mode 2). For two directly coupled resonators, $\kappa_{12} = \kappa_{21}^* = \kappa$ due to conservation of energy. The phase factor of a coupling constant usually does not exhibit physical meaning unless the relative phase is considered [52–54]. When the resonant frequencies of the two resonators are aligned such that $\omega_1 = \omega_2 = \omega_0$, it is also possible to manage the two resonators to realize a $PT$-symmetric system by setting gain-loss balance, or a passive $PT$ system if gain and loss are not balanced. The eigenvalues are given by

$$\lambda_\pm = \omega_0 - i\frac{\gamma_1 + \gamma_2}{4} \pm \sqrt{|\kappa|^2 - \left(\frac{\gamma_1 - \gamma_2}{4}\right)^2}. \qquad (20)$$

If the coupling strength $\kappa$ is larger than a critical value $\kappa_{th} = \frac{|\gamma_1 - \gamma_2|}{4}$, then the two eigenvalues have two real parts and identical imaginary part, corresponding to two supermodes with the same gain/loss feature but split in resonant frequencies. If $\kappa < \kappa_{th}$, however, the two eigenvalues overlap in real frequency, but have different imaginary parts, leading to two optical supermodes with identical resonant frequency but disparate gain/loss. At the phase transition point where $\kappa = \kappa_{th}$, the eigenvalues are degenerate and the associated eigenstates also coalesce, leading to the emergence of an EP. The above mentioned phase transition with changing coupling strength or gain/loss rates define different regimes for $PT$ systems, and can be observed in both optical waveguides [48] and resonators [55].

## 3. Spectral singularity and enhanced sensing

The EP of a non-Hermitian system is characterized with coalescence of both eigenvalues and eigenvectors, which can be engineered by introducing a small perturbation to the system. Generalized perturbation theory on the eigenvalue bifurcation at an $N^{th}$-order EP yields a unique $N^{th}$-root dependence on the perturbation strength [56]. This idea, if implemented in sensors, could benefit the sensing performance by amplifying the sensor response to tiny perturbation according to the $N^{th}$-root law. On the other hand, microcavity sensors based on the silicon photonics provide unprecedented tunability in terms of the system parameters and are ideal candidate for demonstrating EP sensors. We will discuss the theory on the enhancement of response in EP sensors, as well as how EPs are experimentally realized and maintained in these silicon-photonic systems.

### 3.1 Spectral singularity at exceptional points

In the systems with non-Hermitian Hamiltonians, the spectral singularities associated with EPs manifest themselves as the simultaneous coalescence of several eigenvalues and eigenvectors. The Hamiltonian becomes non-diagonalizable at these singularities and can only be transformed into Jordan canonical form by a similarity matrix, where the order of EPs determines the length of Jordan block within the Jordan canonical form. For example, an $N^{th}$-EP corresponds to a Jordan block of length N that takes the form below

$$J_\lambda(N) = \begin{bmatrix} \lambda & 1 & \cdots & 0 \\ 0 & \lambda & & 0 \\ \vdots & & \ddots & \vdots \\ 0 & 0 & & 1 \\ 0 & 0 & \cdots & \lambda \end{bmatrix}. \tag{21}$$

It is of natural instinct to investigate how perturbation to the system will cause bifurcation of eigenvalues and eigenvectors. Square-root and third-root dependence in the bifurcation with respect to the perturbation strength have been found in second- and third-order EPs [57,58], respectively. In a more general case, it has also been shown that the perturbative behavior of a Jordan block of length N follows an $N^{th}$-root relation [56]. Under a small perturbation $\epsilon$ to the elements of Jordan block, the eigenvalue can be expressed in the form of Puiseux series

$$\lambda(\epsilon) = \lambda(0) + \lambda_1 \sqrt[N]{\epsilon} + \mathcal{O}(\sqrt[N]{\epsilon}), \tag{22}$$

which holds as long as the first term $\lambda_1$ is not vanishing. This condition is true for most physical perturbations that could affect the systems. Therefore, it was proposed as a novel method to enhance the response of sensors to small signals [59–65], as the sensor response that is proportional to $N^{th}$-root of a sufficiently small perturbation is much larger than the linear case ($\sqrt[N]{\epsilon} \gg \epsilon$ when $\varepsilon \to 0$). In practice, EPs of lower order are easier to implement and their enhancing effects have been demonstrated in several types of photonic sensors, including nanoparticle sensors [66], gyroscopes [67], thermal sensors [68] and even immuno-assay nano sensors in a plasmonic system [69]. Among them the first two are based on silicon photonics. The great flexibility of parameters in the silicon photonic systems facilitates the realization of EPs and paves a new way to improve sensor performance, which will be discussed in the following sections.

### 3.2 EP-enhanced nanoparticle sensor

The unique $N^{th}$-root topology in the vicinity of an EP can be utilized to enhance the sensor performance, for example, EP enhanced nanoparticle sensing based on an optical microcavity [66]. To understand the amplification of tiny sensor response at EPs, let us consider the conventional case first. When a conventional microcavity is subject to a perturbation induced by some scatterers, like nanoparticles, the Hermitian degeneracy (also known as diabolic point,

DP) between clockwise (CW) and counter-clockwise (CCW) travelling modes is lifted, leading to a mode-splitting in the microcavity [29,70–73]. The strength of mode-splitting at DPs is proportional to the strength of perturbation, which can be derived from the Hamiltonian: Starting from a general scenario — an unperturbed system consisting of a microcavity and $M$ scatterers. The $(M+1)^{th}$ scatterer is introduced as a perturbation, then the total effective Hamiltonian is [59,74]

$$H = \begin{bmatrix} \Omega^{(M)} & A^{(M)} \\ B^{(M)} & \Omega^{(M)} \end{bmatrix} + \begin{bmatrix} V+U & (V-U)e^{-i2m\beta} \\ (V-U)e^{i2m\beta} & V+U \end{bmatrix}. \quad (23)$$

The first term describes the unperturbed system, with the off-diagonal elements $A^{(M)}$ and $B^{(M)}$ representing the intrinsic backscattering. The second term is the Hamiltonian of perturbation. $2V$ and $2U$ are the complex frequency shifts for positive- and negative-parity modes due to the perturbation, $m$ is the azimuthal mode order and $\beta$ is the relative angle of the $(M+1)^{th}$ scatterer. Conventional micocavity sensors working at DPs have no intrinsic backscattering between CW and CCW modes, i.e., $A^{(M)} = B^{(M)} = 0$. Therefore, the complex frequency splitting caused by a perturbation can be computed as $\Delta\Omega_{DP} = 2(V-U) = 2\epsilon$. Here we define $(V-U)$ as the complex perturbation strength $\epsilon$, so we show that the surface of eigenfrequencies for conventional sensors has a linear (cone-shaped) topology around DPs (Fig. 3a).

However, sensors operating at EPs have a completely different topology, which can help enhance sensitivity for small perturbations. Consider the same 2 by 2 Hamiltonian above, if the system is at EP, its backscattering will be fully asymmetric. There is either no backscattering from CCW to CW modes (or vice versa), *i.e.*, $A^{(M)} = 0$ (or $B^{(M)} = 0$). In the case of $B^{(M)} = 0$, for example, the complex frequency splitting for second-order will be

$$\Delta\Omega_{EP} = 2\epsilon\sqrt{1 + \frac{A^{(M)}e^{i2m\beta}}{\epsilon}} \approx 2e^{im\beta}\sqrt{A^{(M)}\epsilon}, \quad (24)$$

which gives rise to a square-root topology (Fig. 3b). As long as the intrinsic backscattering is much larger than the perturbation ($|A^{(M)}| \gg |\epsilon|$), EP sensors will have a greater response compared to conventional DP sensors

$$|\Delta\Omega_{EP}| = 2\sqrt{A^{(M)}\epsilon} > |\Delta\Omega_{DP}| = 2\epsilon. \quad (25)$$

This implies that the sensitivity of sensor to sufficiently small perturbation can be enhanced by operating the sensor at EPs.

EP enhanced nanoparticle sensing has been demonstrated in a silica microcavity-scatterer system [66]. To steer the microcavity to an EP, two silica fiber tips (nano-tips) were used as scatterers. The nano-tips were mounted on two translation stages respectively, which can adjust their positions finely and thus tune the intrinsic backscattering. The EP sensor is realized when the mode-splitting by the first nano-tip disappears in the presence of the second nano-tip, judged by observing unidirectional suppression of reflection signal [30,75]. Then, a third nano-tip was introduced as the target perturbation to characterize the sensor, and the perturbation strength is varied by tuning the position of the

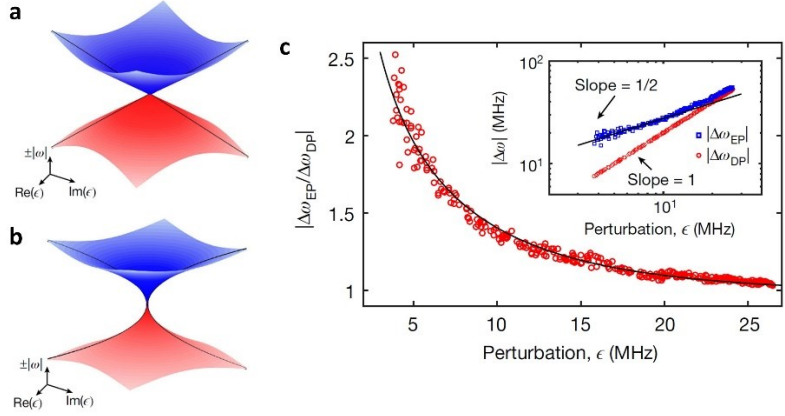

Fig. 3 a, b Surfaces of the eigenfrequencies of (**a**) DP sensors and (**b**) EP sensors. **c** Measured EP enhancement factor upon the variation of perturbation strength. The inset shows the log-log plots of the mode-splitting in the EP and DP sensors. Figures from Ref. [66].

third nano-tip within the optical mode volume. The observed mode-splitting in the EP sensor is larger than that in the DP sensor for any perturbation that is weaker than 25 MHz, and the enhancement factor can go up to 2.5 for perturbation strength less than 5 MHz (Fig. 3c). The inset shows the log-log plots of the mode-splitting as the perturbation strength varies in the EP (blue dots) and DP sensor (red dots). The log-log plot for EP sensor has a slope of 1/2, which validates the existence of a second order EP and its square-root topology.

### 3.3 EP-enhanced gyroscope

The sensitivity enhancing effect at EPs has also been exploited in silicon-photonic gyroscope based on Brillouin laser [67], which consists of a silica disk resonator on a silicon chip coupled to a fiber taper. Brillouin scattering is an important nonlinear process that describes the scattering of photons from phonons. It occurs spontaneously at a low power level, and its stimulated emission excited by a high-power pump can even lead to stimulated Brillouin lasers (SBLs). First- and higher-order SBLs can be excited at a milliwatt pump power, thanks to the ultrahigh optical quality (Q)-factor of the resonator and the fine control of its size [76]. Conventionally (without EPs), when the whole system rotates, the Sagnac-induced frequency shifts for counter-propagating SBLs are opposite, forming a beat note whose beating frequency is proportional to the rotation rate [77,78]. In the EP gyroscope, CW and CCW SBLs on the same cavity mode are excited by sending two pump waves with opposite direction into the resonator (Fig. 4a). To achieve a second-order EP, it is critical to induce dissipative coupling between these two counter-propagating lasing modes, which is achieved by the scattering at fiber taper.

To analyze the EP-enhanced Sagnac effect predicted by previous theoretical work [62,63], we consider the Hamiltonian that governs the time evolution of CW and CCW SBL modes subject to an angular rotation rate of $\Omega$:

$$H = \begin{bmatrix} \omega_0 + \frac{\gamma}{\Gamma}\Delta\Omega_1 & i\kappa \\ i\kappa & \omega_0 + \frac{\gamma}{\Gamma}\Delta\Omega_2 \end{bmatrix} + \begin{bmatrix} -\frac{1}{2}\Delta\omega_{sagnac} & 0 \\ 0 & \frac{1}{2}\Delta\omega_{sagnac} \end{bmatrix}. \quad (26)$$

where $\omega_0$ is the Stokes cavity mode without pump, $\gamma$ is the cavity damping rate, $\Gamma$ is the bandwidth of Brillouin gain, $\kappa$ is the dissipative coupling rate, $\Delta\Omega_j = \omega_{pj} - \omega_s - \Omega_{phonon}$, for $j = 1, 2$, is the frequency mismatch of Brillouin

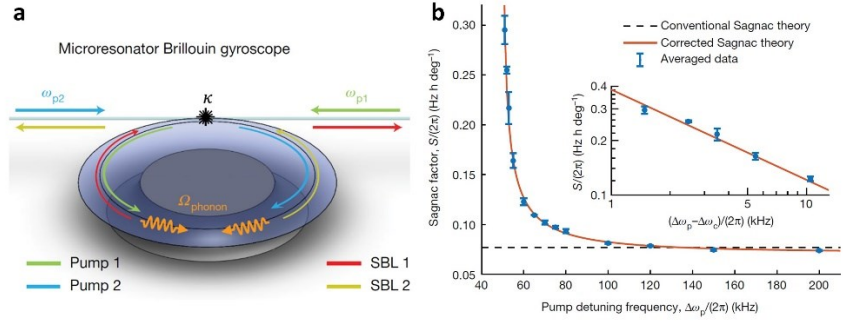

**Fig. 4 a** Schematic of the EP gyroscope. Two SBL modes (*red and yellow arrows*) excited by two pumps (*green and blue arrows*) have dissipative coupling with each other. **b** Experimentally measured Sagnac scale factor as a function of pump detuning (*blue dots*), compared with the theoretical results (*red curve*). The inset is a log-log plot near the EP. Figures from Ref. [67].

scattering corresponding to CW and CCW lasing modes [79], where $\omega_{pj}$ is the pump frequency, $\omega_s$ is the Stokes lasing frequency and $\Omega_{phonon}$ is the Brillouin shift frequency. The second term describes the Sagnac effect that induces a frequency difference between CW and CCW modes $\Delta\omega_{sagnac} = 2\pi D\Omega/(n_g\lambda)$, where $D$ is the resonator diameter, $n_g$ is the group index of the unpumped cavity mode and $\lambda$ is the laser wavelength. The beating frequency is evaluated from the difference of the Hamiltonian's eigenfrequencies

$$\Delta\omega_s = \frac{\gamma/\Gamma}{1+\gamma/\Gamma}\sqrt{(\Delta\omega_p + \Gamma\Delta\omega_{sagnac}/\gamma)^2 - \Delta\omega_c^2}, \qquad (27)$$

where $\Delta\omega_p = \omega_{p2} - \omega_{p1}$ is the pump detuning and $\Delta\omega_c = 2\Gamma\kappa/\gamma$ is the critical pump detuning to reach an EP. To quantify the EP enhancement, the Sagnac scale factor for small rotation rate is defined as

$$S = \left.\frac{\partial\Delta\omega_s}{\partial\Omega}\right|_{\Omega\to 0} = \frac{1}{1+\gamma/\Gamma}\frac{\Delta\omega_p}{\sqrt{\Delta\omega_p^2 - \Delta\omega_c^2}}\frac{2\pi D}{n_g\lambda}. \qquad (28)$$

Therefore, if the gyroscope operates near an EP, i.e., $\Delta\omega_p \to \Delta\omega_c$ in the regime of $|\Delta\omega_p| > \Delta\omega_c$, its small-signal Sagnac scale factor will be much larger than the conventional value $2\pi D/(n_g\lambda)$.

In the experiment, the disk resonator is packaged in a box and experiences a small rotation rate of 410°/h produced by a piezoelectric stage. The measured Sagnac scale factor upon the variation of pump detuning is shown in Fig. 4b. The experiment results (blue dots) match well with the theoretical prediction (red curve), and can be enhanced near EP by a factor of 4 compared to the conventional value (black dash line). Additionally, a log-log plot of five data points near the EP is shown in the inset. The log-log curve has a slope of -1/2, which confirms the existence of $1/\sqrt{\Delta\omega_p^2 - \Delta\omega_c^2}$ factor in the Sagnac scale factor predicted by the theory.

It is also worth noting that EP gyroscope demonstrated here is only exceptional in its large response compared to conventional ones. But the achievement of an exceptional signal-to-noise ratio (SNR) in EP sensors is not straightforward [80,81]. Recently, it has been experimentally demonstrated that the SNR of EP gyroscope is not improved [82], as the enhanced Sagnac scale factor and the Petermann linewidth broadening cancel each other. This is essentially because the non-orthogonality of mode at an EP will lead to a broadening of SBL linewidth. Such enhancement effect is assessed with the Petermann factor [83–86], which poses a fundamental limit to the improvement of EP gyroscope based on SBLs. However, by adopting quantum quadrature measurement scheme or non-reciprocal approach, it is expected that EP amplifying sensors can exhibit enhanced SNR breaking the quantum noise limit [87,88].

## 4. Mode interactions and lasing effects

It is not hard to see that the interaction between optical modes have played critical roles in non-Hermitian photonic devices and systems. Conversely, the non-Hermitian features can exert significant influence on the optical modes and their interactions. As we have discussed above, by coupling optical modes with gain/loss features in confined optical structures, non-Hermitian optical systems render not only complex eigenspectral, but also non-orthogonal eigenmodes, *i.e.*, the eigenstates of the systems are no longer orthogonal. This is true even for *PT*-symmetric systems in unbroken regime where real eigenfrequencies are obtained. What is most interesting is the eigenstate at EP singularity, where two eigenstates not only coalesce, but lead to a special chirality, as will be discussed in this section.

One of the most intriguing results of mode interactions can be found in lasing and emission behavior of non-Hermitian optical systems. Silicon photonic platforms have provided versatile designs of microlasers with high efficiency and low threshold [89]. However, the design of lasers with single-mode operation, controllable emission direction and stability remain challenging, and require the development of novel physics and optical structures. We will examine several counterintuitive phenomena in lasing and emission, which are achieved based on the non-Hermitian mode features and their interactions, taking the advantage of unconventional physical properties such as chirality, non-orthogonality, and variance of eigenspectra around the phase transition points.

### 4.1 Chiral mode at exceptional points

The most exotic phenomena arising from the interaction of optical modes in non-Hermitian systems can be found at EPs. When two eigenstates of a non-Hermitian system coalesce at an EP, one can obtain a self-orthogonal eigenstate which is a superposition of the original basis. For the system comprising of two modes, the eigenstate is typically in the form of $\begin{pmatrix}1\\\pm i\end{pmatrix}$, with a typical $\pm\frac{\pi}{2}$ phase shift between the two components. Such form of eigenstate possesses a particular handedness, which can be defined as chirality [90]. One can visualize the concept of chirality by postulating that the wavefunctions in the original basis consist of two perpendicular polarization directions, and then at an EP, the eigenstate takes the form of a circularly polarized wave and rotates in either CW or CCW direction.

In optical resonators, EPs with specific chiral eigenmodes have been engineered and verified experimentally [91]. Intrinsically, a whispering gallery mode (WGM) microresonator made in silica supports degenerate optical modes propagating along the circular boundary of the device in both CW and CCW directions. A nanotip made by etching a fiber-taper end into a cone shape is positioned close to the rim of the resonator and can perturb the evanescent field, inducing backscattering (with strength $\epsilon_1$) of the optical field and coupling the CW and CCW optical modes. The system under perturbation supports symmetric and antisymmetric standing wave supermodes, each constituting CW and CCW components. A second nanotip separated from the first by an azimuthal distance $\beta$ can be exploited to break the chiral symmetry of the optical structure by introducing an additional perturbation $\epsilon_2$ to the evanescent field. The non-Hermitian Hamiltonian of the system under the perturbation of the two nanoscatterers can be derived [59,74]

$$H = \begin{pmatrix} \omega_0 - i\frac{\gamma_0}{2} + \epsilon_1 + \epsilon_2 & \epsilon_1 + \epsilon_2 e^{-2im\beta} \\ \epsilon_1 + \epsilon_2 e^{2im\beta} & \omega_0 - i\frac{\gamma_0}{2} + \epsilon_1 + \epsilon_2 \end{pmatrix}, \qquad (29)$$

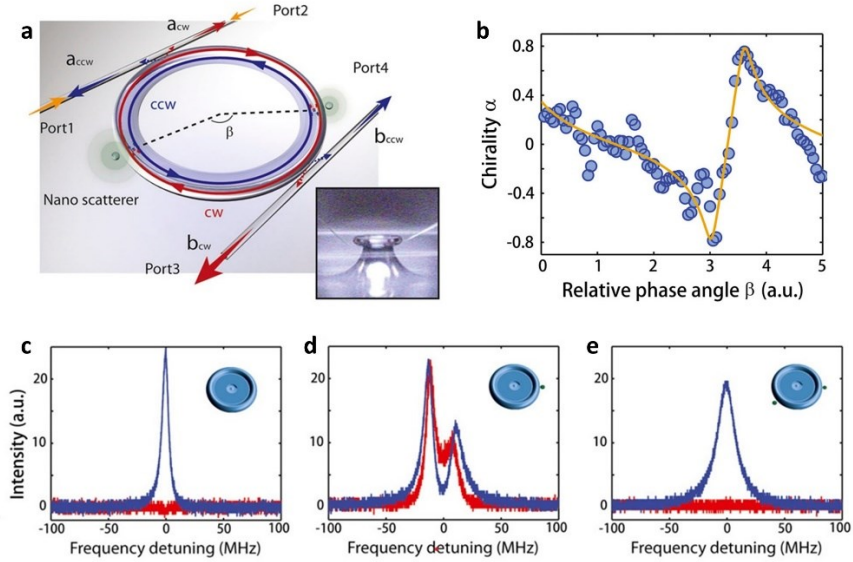

**Fig. 5** Chiral modes and unidirectional lasing at an EP. **a** Schematic diagram for realizing chiral modes at EPs in a silica microresonator. The resonator is coupled to two fiber-taper waveguides and perturbed by two silica nanotips which can adjust the backscattering between CW and CCW modes. **b** Measured chirality as a function of the relative phase angle between the two nanotips. **c-e** Measured light intensity from port 3 (red) and port 4 (blue), for (**c**) no nanotip perturbation, (**d**) one-nanotip perturbation, and (**e**) an EP achieved by two-nanotip perturbation. Figures from Ref. [91].

where the CW and CCW modes with azimuthal mode number $m$ have identical resonant frequency $\omega_0$ and intrinsic loss rate $\gamma_0$. The supermodes are given by $\psi_\pm = \sqrt{A}\psi_{ccw} \pm \sqrt{B}\psi_{cw}$, where $A = \epsilon_1 + \epsilon_2 e^{-2im\beta}$ and $B = \epsilon_1 + \epsilon_2 e^{2im\beta}$. By manipulating the position of the second nanotip, one can engineer one of the non-diagonal element ($A$ or $B$) to vanish, leading to an EP with a singular form of $H$ (one non-diagonal element vanishes) and only one eigenstate $\psi_{EP} = \sqrt{A}\psi_{ccw}$ or $\psi_{EP} = \sqrt{B}\psi_{cw}$. Such eigenstate is purely chiral with one directional propagation, by which we can define chirality -1 for a CCW eigenmode and 1 for a CW eigenmode [91]. By tuning the relative phase angle β, one can change the value of chirality continuously and periodically between -1 and 1, due to the periodic variation in the phase of the non-diagonal coupling element in the Hamiltonian (Fig. 5b).

### 4.2 Unidirectional lasing

The chirality at an EP can manifest itself with the assistance of optical modes operating beyond lasing threshold and the directionality in laser emission. The chiral mode at an EP which rotates in one direction has been demonstrated in a silica microresonator with gain and unidirectional lasing behavior is observed [91,92]. The gain medium is introduced by spin coating sol-gel solutions containing $Er^{3+}$ ions to a silicon wafer and forming a layer of silica material with about 2 $\mu m$ thickness by thermal annealing [35,93–95]. The silica microtoroid is then made in the standard photolithography, wet and dry etching, and reflow procedures. In experiments, the device is characterized by coupling to double fiber-taper waveguides with the configuration shown in Fig. 5a. By injecting pump light in the 1450 nm band from port 1, the device can lase around the wavelength of 1550 nm with detectable emission of laser from port 3 (Fig. 5c). Such microlaser can emit bidirectional laser collected by both port 3 and port 4 once it is perturbed by a nanotip, because the scatterer induces coupling between the CW and CCW propagating fields (Fig. 5d). However, if a second nanotip is applied and the resonator is tuned to an EP where backscattering from CW to CCW directions vanishes, then the only surviving eigenstate is in the CW direction, and thereby the laser emission is only measured from port 3 (Fig. 5e). Further adjusting the second scatterer to engineer another EP with CCW eigenmode and reversing the chirality will lead to unidirectional lasing in only the CCW direction.

The chiral mode can also be exploited to generate vortex beam and orbital angular momentum lasers if the amplifying mode emit vertically out of the plane of the circular optical path in the cavity [96]. In addition, some behavior opposite to what is described in this part has also been reported. For instance, in a system by a precise design of backscattering ratio between CW and CCW modes, spontaneous emission of emitters can be coupled to a Jordan vector of the EP microcavity instead of the channel aligned with the eigenstate [97].

### 4.3 Single mode laser

Cavities in lasers, which provide wavelength selection and coherent amplification of light, usually support multiple resonances and optical modes [98]. Due to the homogeneous broadening of gain spectrum of the active medium [99], adding gain to the desired mode is often accompanied by generating lasing in the neighboring modes, giving rise to mode competition and laser instability. Therefore, to achieve single-mode operation for lasers is one of the critical tasks in the design and management of lasing systems.

By leveraging the threshold of $PT$-symmetry breaking, single-mode lasing operation can be established in microcavities which support multiple modes in the gain spectrum [100,101]. The system is composed of coupled active and passive microresonators with identical geometry, thus with the same resonant frequencies in the spectra. Due to the inhomogeneously broadened gain profile, the different modes in the active resonator will experience different amount of gain. The strong coupling between the optical modes in the two resonators will cause frequency split, generating pairs of modes in the eigenspectra of the whole system. The maximum gain is first tuned to be equal to the loss in the other resonator, so that the system stays in the $PT$ unbroken phase, where each pair of mode will remain non-amplifying, due to the fact that the two supermodes have the same imaginary part of the eigenfrequency and remains below lasing threshold. By increasing the pump power and the gain in the active resonator, one pair of mode with the largest gain will reach the $PT$-transition threshold and enter the $PT$-broken regime, bringing one supermode with amplification and the other with dissipation. Therefore, only the supermode with net gain will lase, while other supermodes in other mode pairs still remain in the $PT$-unbroken regime and stay below the lasing threshold.

Such lasers have been realized in various semiconductor platforms [100,101], and may bring novel routes of laser management and control in silicon photonics.

### 4.4 Revival of lasing by loss

More unconventional lasing behavior can be engineered by leveraging non-Hermiticity of open lasing systems, especially around the EPs where phase transition occurs. One counterintuitive example can be found in a photonic molecule comprising coupled microresonators, where increasing the loss rate of one microresonator can induce the suppression and revival of laser emission in another, seemingly contradictory to the common sense that loss always acts as a negative and detrimental factor to laser operations [36].

In the demonstration of this idea, two silica microtoroid resonators are coupled via evanescent field and probed by fiber-taper waveguides (Fig. 6a). One resonator is active with the assistance of Raman gain. The Raman nonlinear effect which is commonly found in silica material induces frequency shift of the pump light, and, with the help of a cavity, can induce coherent amplification of the Raman signal light, leading to stimulated Raman emission, i.e., the Raman lasing [89]. The other resonator is purely lossy and subject to external loss exerted by a chromium (Cr) absorber. The Raman laser in this scenario is thus embedded in a non-Hermitian context with the interaction between optical gain and loss units, and will undergo unconventional enhancement and reduction with the change of the external loss on one of the resonators. When the loss rate of the second resonator is increased, the Raman lasing threshold first increases corresponding to a suppression of lasing, and then decreases, leading to revival of laser (Fig. 6b, c). The change from laser suppression to laser revival occurs around the phase transition point, i.e., EP. One can understand the phenomenon based on the field distribution in the two resonators. When the loss in the second resonator is relatively small, the system supports two supermodes with identical loss rates but different resonances, and the energy distribution in the two resonators is symmetric. With more loss introduced into the lossy resonator by the Cr absorber, the system will undergo a phase transition passing the EP and enter the broken regime where the supermodes with amplification and dissipation will be localized in the active and the lossy resonator, respectively. The higher contrast of gain and loss between the two units intensifies the mode localization. Therefore, when the external loss increases, the field intensity in the lossy resonator becomes even smaller while that in the active resonator is enlarged. Such redistribution of the optical energy among the system vividly presents the process where the field in a lossy unit sacrifices for the survival of the active counterpart as the imbalance of dissipation among the system is amplified.

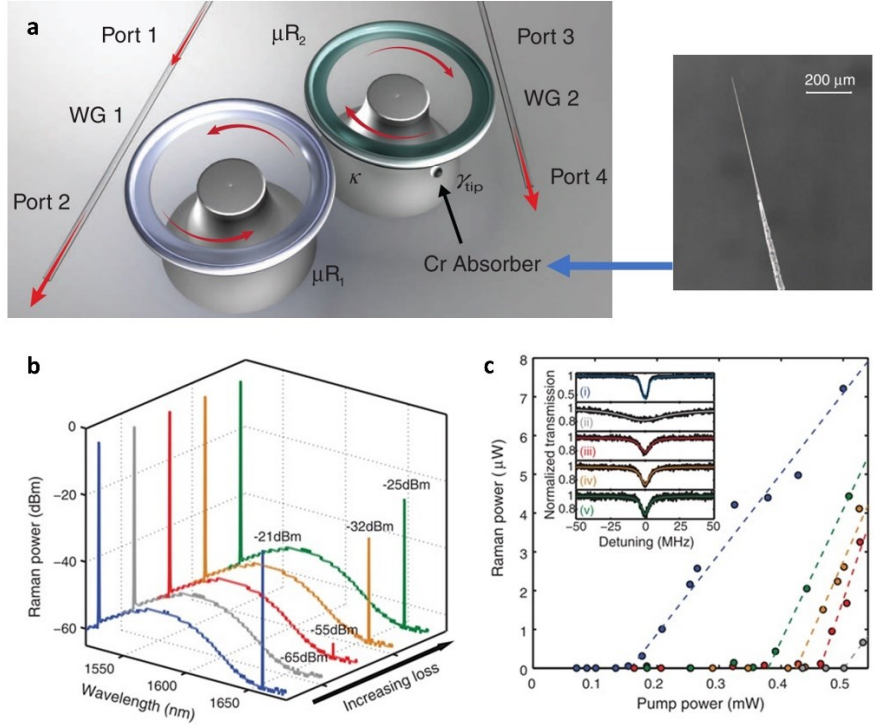

**Fig. 6** Revival of Raman laser by external loss. **a** Schematic diagram of the coupled microtoroid resonators $\mu R_1$ and $\mu R_2$ coupled to two waveguides. A Chromium (Cr) absorber (inset on the right) is used to tune the loss of $\mu R_2$. **b** Raman lasing spectra when the loss of $\mu R_2$ is increased by adjusting the position of the Cr absorber. **c** Variation of Raman laser power with the pump power for different loss of $\mu R_2$. The inset shows the transmission spectra near the pump frequency with loss increase from top to bottom. Each color in (**a**) and (**b**) corresponds to a value of induced loss on $\mu R_2$. Figures from Ref. [36].

Such bizarre lasing behavior managed by the loss is a direct result of the interaction between optical modes with gain and loss, as well as the redistribution of optical field among the whole system with the occurrence of phase transition. Similar observation has also been made in other optical structures and platforms [50,102].

### 4.5 Peterman factor and laser linewidth

The linewidths of lasers determined by quantum noise can be significantly influenced by modal coupling in laser systems. With the interaction between the optical modes with different gain/loss features, non-Hermitian systems render non-orthogonal eigenmodes and consequently have nontrivial effects on the laser linewidths. It has been predicted and demonstrated that the non-orthogonality of the eigenmodes can lead to amplified quantum noise in lasers, as well as significant enhancement of laser linewidth quantified by the Peterman factor beyond the Schawlow-Townes quantum limit [83–86,103]. Moreover, at an EP where eigenmodes are self-orthogonal, the laser linewidth is predicted to be extremely broadened [104].

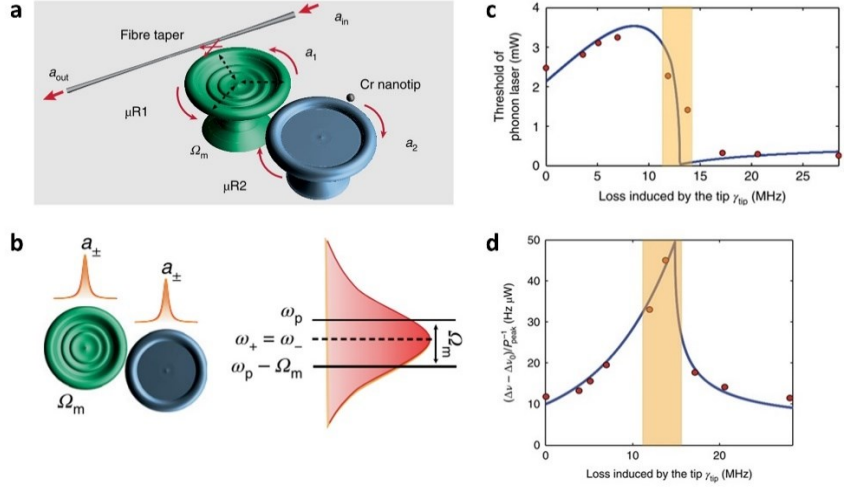

**Fig. 7** A phonon laser at an EP. **a** Schematic diagram for coupled microtoroid resonators supporting WGMs with field amplitudes $a_1$ and $a_2$ coupled to a fiber taper waveguide. The green resonator $\mu R_1$ supports mechanical oscillation with resonant frequency $\Omega_m$. The blue resonator $\mu R_2$ is perturbed by a Cr nanotip so that the optical loss rate can be tuned externally. **b** Optical supermode and spectrum when the system is at the EP. **c** The measured threshold of the phonon laser varying with the external loss induced by the nanotip. **d** The measured phonon laser linewidth as a function of the external loss induced by the nanotip. Figures from Ref. [105].

Such effect has been investigated for a phonon laser operating at an EP realized in a silica microresonators [105]. Microtoroid structures with silica microdisks supported by a silicon pillars can support mechanical oscillations, which can be excited by radiation pressure induced by intracavity optical fields [106]. A phonon laser, which arises due to the coherent amplification of a mechanical mode, can be realized if the mechanical gain is provided by the two optical supermodes with split frequencies that form a two-level structure [107]. If the split eigenspectrum has certain linewidths covering a range that can trigger the mechanical oscillation supported by the microtoroid structure, then the two-level system acts as a gain medium and interact coherently with the mechanical mode, thus building a phonon laser scheme (Fig. 7a). In such a phonon laser system, an EP for the optical modes can be engineered by tuning the optical loss rate of one resonator, and subsequently lead to intriguing phonon lasing behavior. When increasing the loss of a resonator externally by a Cr nanotip, the phonon lasing threshold is first lifted and then falls abruptly around the EP (Fig. 7b). The linewidth of the phonon laser follows a similar trend, i.e., it first increases and then rapidly decreases near the EP, before finally arising slowly (Fig. 7). The broadening of the linewidth around the EP verifies the excessive noise in the optical gain medium resulting from the strongly non-orthogonal eigenstates.

The enhanced noise in EP laser systems has significant influence on the performance of EP sensors which rely on enhanced lasing mode splitting. It has been shown in a high-$Q$ silica microresonator that the ring-laser serving as a gyroscope does not offer enhanced signal-to-noise ratio compared to conventional ones, as a result of the mode non-orthogonality and Petermann-factored noise [82]. However, breaking the conventional quantum noise limit may be possible with the help of quadrature detection scheme [88].

### 4.6 Other non-Hermitian lasing behavior

The inverse operation of laser, which is referred to coherent perfect absorber (CPA), can also exhibit peculiar features in non-Hermitian setting. It is proposed and demonstrated that a *PT*-symmetric waveguide with gain/loss modulation operating in broken phase can simultaneously function as a laser and a CPA [108,109]. Furthermore, the degenerate CPA solutions coalescing to an EP can exhibit singular lineshape in absorption spectrum [110] and chiral absorbing feature [111].

Phonon laser operation is also investigated theoretically for an *PT*-symmetric optical medium [112], where EPs are found to enhance the intracavity photon number and thus the optical pressure, leading to lower threshold in phonon lasing. Moreover, it is proposed that optical amplifiers operating at EPs can show a better gain-bandwidth scaling [113]. In addition, EPs are also predicted to be able to control and turn off coherent emission above the lasing threshold in laser systems [114]. We expect a near-future demonstration of these predictions in the silicon photonic platforms.

## 5. Scattering properties and light propagation

Scattering properties of light describe the relation between output and input optical signals in optical devices and systems, laying the foundation of a great many applications in optical communication and information processing. Optical devices in silicon and silica material such as optical fibers, on-chip waveguides, microresonators and photonic lattices have played indispensable roles in guiding light transport due to their transparency and low material loss. Furthermore, silicon/silica photonic structures allow for strong nonlinear optical effects [115–117] such as Kerr effect, optothermal effect, Raman scattering, optomechanics [106], *etc.*, which significantly influence on optical propagation and dynamics. By leveraging EPs and $PT$-symmetry in optical structures, several unconventional scattering properties emerge such as unidirectional reflection and modulated electromagnetically induced transparency (EIT). In addition, the integration of $PT$-symmetry with nonlinear optical effect [13] enables novel methods of designing nonreciprocal light transport with high performance.

### 5.1 Unidirectional reflection at exceptional points

We consider a one-dimensional $PT$-symmetric photonic heterostructure, where complex index modulation yields spatial separation of gain and loss regions. The relation between the output and input electromagnetic waves can be described by a scattering ($S$) matrix as

$$\begin{pmatrix} a_{out,L} \\ a_{out,R} \end{pmatrix} = S \begin{pmatrix} a_{in,L} \\ a_{in,R} \end{pmatrix} = \begin{pmatrix} r_L & t \\ t & r_R \end{pmatrix} \begin{pmatrix} a_{in,L} \\ a_{in,R} \end{pmatrix}, \quad (30)$$

where $a_{in,L(R)}$ represents the input optical wave from the left (right), and $a_{out,L(R)}$ is the output optical wave from the left (right). Generalized unitary relation leads to the conservation relation [118]

$$|T - 1| = \sqrt{R_L R_R}, \quad (31)$$

where $T = |t|^2$, $R_L = |r_L|^2$ and $R_R = |r_R|^2$. Specifically, the transmission rate $T$ is smaller than 1 when the system is in the $PT$-symmteric unbroken regime, and larger than 1 when the $PT$-symmetric phase is broken. Peculiar phenomena in the scattering can happen at the phase transition point, *i.e.*, the EP, in the $PT$-symmetric 1D scattering potential, where the eigenspectra turned from real to complex and the transmission $T$ is equal to 1. As a result, the vanishing left hand side of the conservation relation yields a zero reflection rate for either $R_L$ or $R_R$, leading to unidirectional reflection of incoming optical field [119]. The theory can also be extended to passive $PT$-symmetric systems.

Such phenomenon can be observed in synthetic silicon photonic structure with a modulation of complex refractive index along the light propagation direction. For instance, Feng et al. demonstrated a unidirectional reflectionless Bragg grating structure based on a Si waveguide that is embedded inside SiO$_2$ (Fig. 8a) [120]. The realization of a passive $PT$-symmetric potential is achieved by introducing a periodic modulation of dielectric permittivity in the $z$ direction $\Delta \varepsilon = cos(qz) - i\delta sin(qz)$, where $q = 2k_1$ and $4n\pi/q + \pi/q \leq z \leq 4n\pi/q + 2\pi/q$, with $k_1$ being the wavevector of the fundamental mode. EP occurs when the parameter $\delta$ is equal to 1. The change of the real and imaginary parts of the refractive index are respectively introduced by depositing Si and germanium (Ge)/chrome (Cr) bilayer combo structures on the Si waveguide. An on-chip waveguide directional coupler made of Si is fabricated along with the $PT$-symmetric waveguide to measure the reflection rates from both forward and backward directions. The measured forward reflection is much larger than the backward reflection over a broad telecom band (Fig. 8b), verifying the unidirectional reflectionless light transport at an EP.

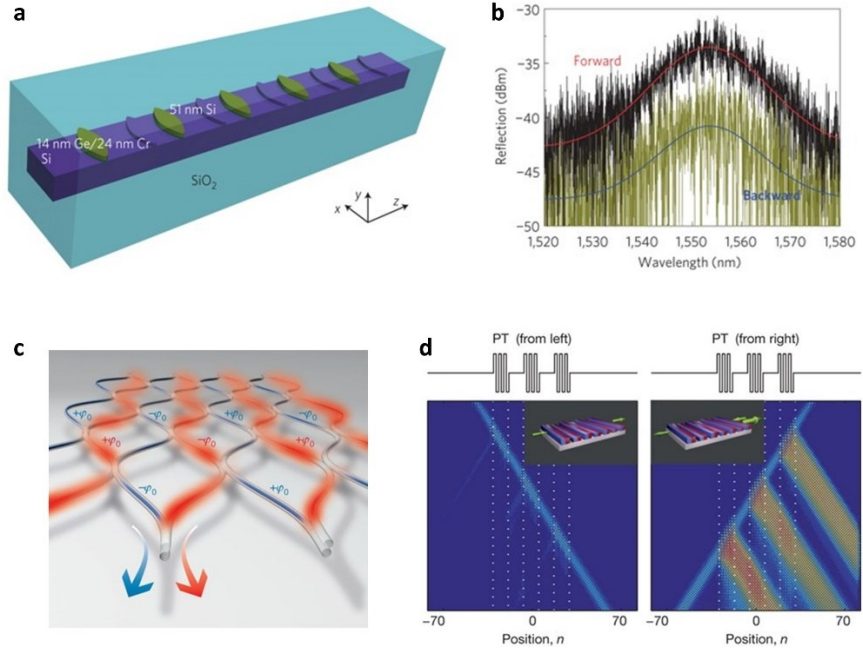

**Fig. 8** Unidirectional reflectionless light propagation at exceptional points. **a** A passive ***PT***-symmteric Bragg structure made of a 800-nm-wide Si waveguide embedded in SiO$_2$ with periodic modulation of the complex refractive index engineered by periodic implementation of Ge/Cr and Si layers. **b** Measured reflection spectrum for the device in (**a**) from both forward and backward directions. **c** ***PT***-symmetric photonic lattice made of periodic layout of fiber loops with gain (red) and loss (blue). The phase shift $\pm\varphi_0$ impose a symmetric real part of the potential. **d** Probing the grating at an EP by left and right incoming light beams. The structure is invisible from left, but visible from right. Color scale: logarithm, red for high intensity and blue for low intensity. Figures **a** and **b** from Ref. [120], and figures **c** and **d** from Ref. [123].

As reflection of light is generally generated by scattering of an optical structure or system, the suppression of reflection can make an object invisible. Conventionally, invisibility can be engineered by encompassing an object with a cloak medium. Here the suppression of reflection from one side at EPs can be exploited to create optical structures with complete unidirectional invisibility over a broad frequency range [121,122]. Experimental demonstration has shown that a temporal *PT*-symmteric photonic lattice consisting of optical fiber loops can be made invisible from one side (Fig. 8c, d) [123]. Furthermore, the phenomenon is found to be robust against Kerr nonlinearity and perturbations [121], and have been proposed in two-layer slab structures [124], one-dimensional photonic crystal [125,126] and various other platforms [127].

### 5.2 Nonreciprocal light transport in nonlinear parity-time symmetric systems

Nonreciprocal light propagation breaks the symmetry in transmission for opposite direction of light propagation, allowing light to propagate only in one direction and completely blocking the opposite transmission [128]. Nonreciprocal phenomena have found widespread adoption in photonic applications such as building isolators protecting lasers from the damaging effect of reflection signals, as well as designing circulators that route directional light propagation among different ports. To realize nonreciprocal light transport, the Lorentz reciprocity must be broken [129], typically by magneto-optic effect, optical nonlinearity, or temporal modulation of materials. As the magneto-optical effect is weak in many materials, bulky structures are needed which brings difficulty to the integration of nonreciprocal devices. As a result, nonlinear [130–134] and time-dependent [135–137] effects are paid much attention in the attempt to break Lorentz reciprocity.

The nonreciprocal light transport is observed in a *PT*-symmetric nonlinear system as a result of the strongly enhanced nonlinearity in the *PT*-broken regime [55]. The system is composed of coupled WGM microtoroid resonators made of silica with different gain/loss features and two fiber-taper waveguides as input and output channels. Resonators are fabricated on the edges of two substrates so that the coupling strength between them can be adjusted by manipulating their separation via nano positioners (Fig. 9a, b). The gain is introduced to the first resonator by $Er^{3+}$ ion dopants and tuned by the pump light so that the net gain balances the intrinsic loss in the second resonator. The *PT*-phase transition can be observed by varying the coupling strength via adjusting the gap between the two resonators. Nonlinear effects such as gain saturation and Kerr effects can happen in silica microresonators and can be triggered by large intracavity optical intensity. In the *PT* unbroken regime, the system has two supermodes without amplification or dissipation, and thus exhibits linear response and reciprocity in the transmission when probed by small power signals (Fig. 9c). However, in the *PT*-broken regime, the system supports two supermodes with amplification and dissipation which are localized in the active and passive resonators, respectively, regardless of the input direction. As a result, the nonlinearity in the active resonator is significantly enhanced by the localized optical field, giving rise to strong nonreciprocity. A strong contrast occurs in the forward and backward transmission where the output signals are collected from the lossy and active resonator sides, respectively (Fig. 9d). Moreover, the complete suppression of the forward transmission is observed, with the threshold of the nonreciprocal light transport as low as $1\mu W$.

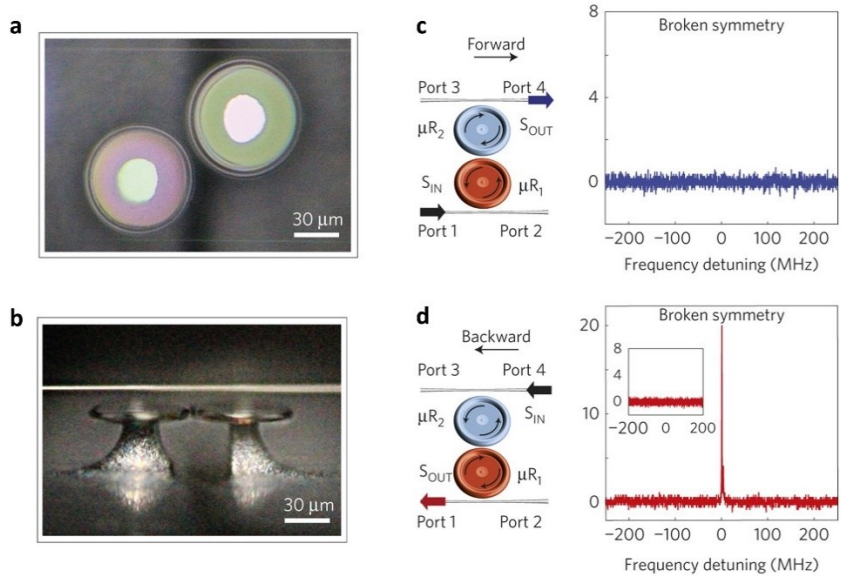

**Fig. 9** Nonreciprocal light propagation in nonlinear PT-symmetric photonic molecules. **a,b** Top view (**a**) and side view (**b**) of the coupled silica microtoroid resonators with gain and loss. **c** Schematic for measuring the forward transmission and the resulting spectrum in the PT-broken regime. **d** Schematic for measuring the backward transmission and the resulting spectrum in the PT-broken regime. Figure from Ref. [55].

Due to the giant nonlinearity enhanced by gain medium and *PT*-phase breaking [138], the *PT*-symmetric resonator systems which operate near the boundary of stability is an ideal platform to study versatile nonlinear static or dynamic behavior. The rich tuning degrees of freedom such as waveguide-resonator coupling, probe power and frequency detuning, make it possible to engineer actively controlled optical isolators [139,140]. The nonreciprocal wave transport enabled by *PT*-symmetry has been explored in mechanical [138], acoustic [141] and electronic [15] systems as well.

### 5.3　Electromagnetically induced transparency in non-Hermitian systems

Electromagnetically induce transparency (EIT) describes the phenomenon that light can pass through an opaque dielectric medium due to the destructive interference established by strong coherent light-matter interaction. The strong cancellation of absorption enabled by EIT leads to a vast change of material dispersion which gives rise to slowing down of the group velocity of light [142–144]. Such slow light behavior is plays an indispensable role in optical memory and storage [145]. EIT was first proposed and demonstrated in atomic systems that support Λ-type energy levels [146,147], where the absorption of probe light resonant with one transition can be eliminated due to the presence of the strong coupling light interacting with the other electronic transition. Optical analogue of EIT have been studied in various linear optical systems [148–152], with the advantage of on-chip integration and room temperature operation. Typically, coupled optical elements are needed in order to mediate photon absorption via destructive interference of optical fields. For example, coupled microresonators with contrasted quality ($Q$) factors can be utilized to generate a Λ-type three level system, and allow for EIT operation without the need of strong coherent coupling light [153,154]. Furthermore, optomechanically induce transparency (OMIT) can be found in optomechanical cavities thanks to the interaction between red-detuned pump light, probe light on resonance with the cavity, and the mechanical oscillation [155–157]. As we will see, the non-Hermitian properties of the system can offer large degrees of freedom for technical implementation and operational capability in EIT study.

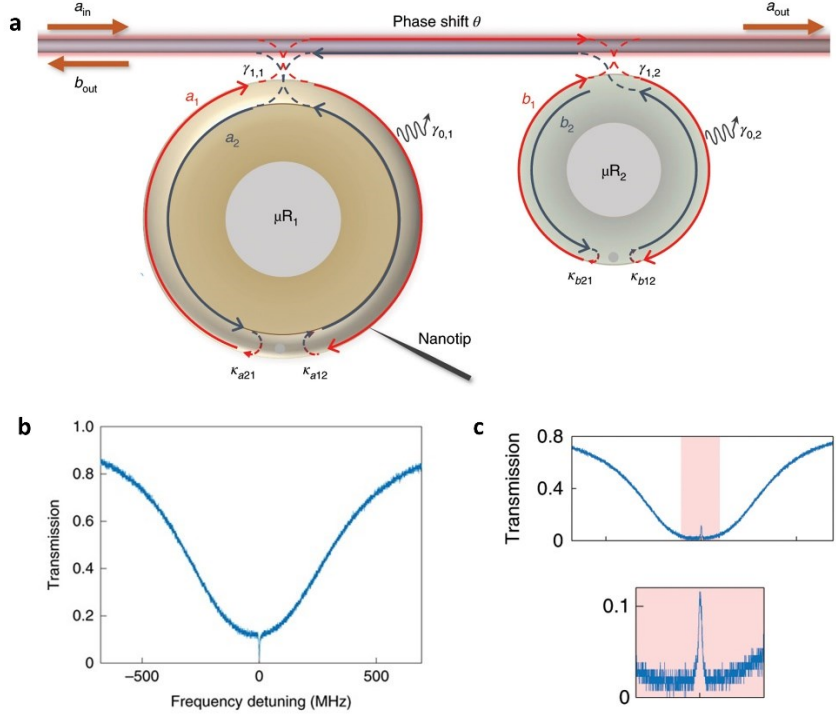

**Fig. 10** Electromagnetically induced transparency (EIT) controlled by chiral modes at exceptional points. **a** Indirectly coupled microresonators for realizing the optical analogue of EIT. By tuning the phase $\theta$, destructive interference can happen in the optical path loop established by backscattering on the two resonators. $\mu R_1$ has a smaller optical loss than $\mu R_2$, and is perturbed by a nanotip so that its chirality can be adjusted and exceptional points can be realized. **b** Transmission spectrum under the condition that $\mu R_1$ is at an exceptional point with chirality -1. **c** Transmission spectrum under the condition that $\mu R_1$ is at an exceptional point with chirality 1. The inset on the lower panel shows a close up of the central transparency window. Figure from Ref. [158].

EPs have been found to offer novel manner to control EIT in non-Hermitian silica optical structure. The eigenmode at EPs with certain chirality has been exploited to realize switching on/off EIT in an indirectly coupled WGM microresonator system [158]. The two microtoroid resonators with different $Q$ factors are coupled to the same fiber-taper waveguide and probed by injecting single mode laser with continuous wavelength scanning (Fig. 10a). In the ideal situation where no scatterer exists on the two resonators, they are not coupled to each other, due to the fact that light can propagate only in one direction so that any light coming out of the first resonator cannot return back into the cavity mode. Nevertheless, backscattering on resonators are ubiquitous due to particle accumulation and surface roughness. Therefore, optical field in each resonator can be reflected from CW into CCW direction (or inversely) so that an optical path loop is formed in the system. Depending on the phase accumulation on the loop which can be changed by the distance between the two resonators, destructive interference can occur, leading to suppression of cavity field and transparency in the forward transmission spectrum. The transparency window exhibits splitting due to the lift of degeneracy in the resonator with higher $Q$ factor. By an additional nanotip, one can tune $\mu R_1$ to an EP with a chirality of -1 for CCW eigenmode or 1 for CW eigenmode, which affect EIT very differently.

For a chirality of -1, the reflection from CCW to CW vanishes, and the optical path loop is broken. Thus, EIT is switched off and a narrow absorption dip is shown in the transmission spectrum (Fig. 10b). By adjusting the position of the nanotip and reversing the chirality to 1, the reflection from CCW to CW become present and destructive interference can occur with precise tuning of the distance between the two resonators. As a result of the degenerate eigenmode at the EP, a narrow transparency window appears in the transmission spectrum, leading to a standard EIT phenomenon (Fig. 10c). Therefore, this scheme provides a method of leveraging non-Hermitian eigenstates for the control of all-optical analogue of EIT in optical resonators.

Non-Hermitian features have also been proposed to modify OMIT processes in cavity optomechanical systems. For example, OMIT can be engineered in a WGM resonator with its chirality being adjusted by two scatterers [159]. By adjusting the position of one scatter, one can change the chirality of the optical field which modifies the group velocity of light and enable switching between slow and fast light. In a system of coupled active and passive resonator, the requirement of the driving amplitude and the optomechanical coupling strength for OMIT can be greatly relaxed due to the fact that the intracavity optical field is enhanced by the presence of optical gain [160]. The non-Hermitian properties thus offer a wealth of new avenues in the realization of dispersion modification, slow light manipulation and optical storage [161,162].

## 6. Topological features and mode switching

The special topological features of EPs not only attract attention in the applications of sensors, but also unveil profound physics in the dynamical behavior of eigenstates under variation of parameters. For example, the neighborhood of a second order EP in the parameter space was found to resemble the Riemann surface of the square root. Any cyclic parametric loop around the EP has a $4\pi$ periodicity, and adiabatic evolution along such loop is predicted to have many interesting effects, such as state transfer [163–165] and geometric phase [166]. More significantly, it was also shown in both numerical and theoretical methods, that anti-adiabatic jump is bound to happen. No matter how smooth the evolution is, there will always be a breakdown of adiabaticity at least for one mode, leading to a chiral exchange of modes. That is, the final state is only determined by the encircling direction in parameter space. Experimental demonstration of this phenomenon has been done in several systems, including a silicon-waveguide system where the time evolution of parameters can be conveniently implemented. Asymmetric mode switching by encircling EPs paves new way to manipulate the evolution of modes in non-Hermitian systems.

### 6.1 Dynamics of encircling EPs

An adiabatic process, in quantum mechanism, is a process that allows the system to stay in the instantaneous eigenstates when the variation of time-dependent Hamiltonian $H(t)$ is sufficiently slow. As the quantum adiabatic theorem states [167,168], if the system is initially in the n-th eigenstate of the initial Hamiltonian $H(0)$, it will still be in the n-th eigenstate of $H(t)$ at any given time $t$ in the adiabatic process, only picking up some phase factors. In a more general case, the Hamiltonian is defined by some parameters $\lambda(t)$ that evolves with a time period of $T$, that is, $\lambda(0) = \lambda(T)$, and the state of system will acquire an additional phase (Berry phase) that is connected to the geometric property of the closed loop [169–172].

In the context of non-Hermitian Hamiltonian, the study of adiabatic processes predicts new phenomena. In the vicinity of a second-order EP associated with a non-Hermitian Hamiltonian, the parameter space has a self-intersecting topology with a $4\pi$ periodicity [173]. This can be shown using Puiseux series to express the two orthogonal states near an EP [5],

$$\psi_\pm(\lambda) \approx A_\pm(\psi_{EP} \pm \sqrt{\lambda}\psi_1), \qquad (32)$$

where $\lambda$ is the parameter and EP is defined at the origin. To satisfy the orthogonality and unitary normalization conditions, the amplitudes are evaluated as $A_+ = -iA_- = \left[2\sqrt{\lambda}\langle\psi_1|\psi_{EP}\rangle\right]^{-1/2}$. It seems straightforward that under stationary condition, encircling an EP (choose a circle of radius $R$ around the EP, that is, $\lambda = Re^{i\varphi}$) would result in an exchange of eigenstates and acquisition of Berry phase [166],

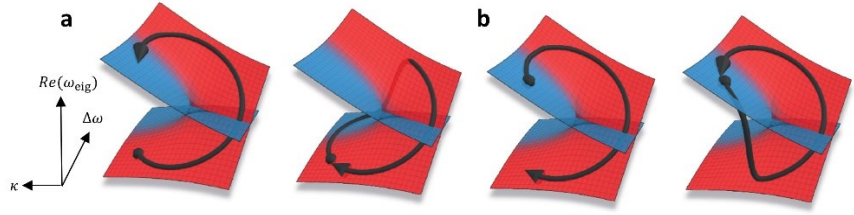

**Fig. 11** Encircling a second-order EP starting from the state on (**a**) lower or (**b**) upper sheets. Due to the breakdown of adiabaticity, encircling the EP in both directions yields a chiral switching of modes. Figures from Ref. [179].

$$\begin{aligned}\psi_\pm\big(\lambda(\varphi=2\pi)\big)&=\mp\psi_\mp\big(\lambda(\varphi=0)\big),\\ \psi_\pm\big(\lambda(\varphi=4\pi)\big)&=e^{i\pi}\psi_\pm\big(\lambda(\varphi=0)\big),\end{aligned} \qquad (33)$$

which has been verified experimentally in various systems [174,175]. Further work, however, found that the small non-adiabatic coupling would inevitably lead to an abrupt transition between the states, preventing adiabatic mode switching to be accomplished in both directions of the closed loop [176–178].

Taking a non-Hermitian two-mode system for example, its dynamics is described by a Schrödinger-type equation $i\,d\psi/dt = H\psi$, where $\psi$ is the amplitude of the modes. The system has a Hamiltonian in the form below

$$H = \begin{bmatrix} \omega_1 - i\dfrac{\gamma_1}{2} & \kappa \\ \kappa & \omega_2 - i\dfrac{\gamma_2}{2} \end{bmatrix}, \qquad (34)$$

where $\omega_j$ and $\gamma_j$, for $j=1,2$, are the resonant frequencies and loss rates of the modes, respectively, and $\kappa$ is the coupling rate. Then the parameter conditions for an EP are $\Delta\omega \equiv \omega_2 - \omega_1 = 0$ and $\kappa = |\gamma_2 - \gamma_1|/4$. The surface of eigenfrequencies with respect to $\Delta\omega$ and $\kappa$ has two Riemann sheets, on which dynamic encircling the EP with different initial states and directions is considered. If the initial state is on the red sheet (Fig. 11a), counter-clockwise encircling the EP can be overall adiabatic and yield a state exchange, while the clockwise one experiences a breakdown of adiabaticity and goes back to the original state; similar phenomenon also happens when the initial state is on the blue sheet (Fig. 11b). Therefore, encircling an EP gives rise to a unique chirality that the final state is determined by the encircling direction, regardless of the initial state.

The chiral exchange of states by encircling EPs was first demonstrated in two kinds of platform. The first one is based on the metallic microwave waveguide [179], in which spatial deformation of waveguide and a special absorption region are designed to steer the system around its EP. The second experiment is conducted in an optomechanical system [180], where the mechanical EP of a membrane inside a cryogenic optical cavity is enclosed by changing laser parameters. These experiment not only inspire more theoretical work and application proposals [181–185], but also provide practical ideas to precisely vary system parameters while maintaining the position of EPs. On the other hand, silicon photonics offer huge flexibility in the tuning of parameters and thus is also a promising candidate in the realization of topological operation around EPs. As we will discuss in the next section, demonstration of chiral mode switching can be done in a silicon platform fabricated by a single lithography, without the need of any cryogenic condition or complicated absorption region.

## 6.2 Asymmetric mode switching

The demonstration of chiral exchange of states by encircling EPs relies on precise control of system parameters, which was also achieved in the silicon-based waveguides [186]. The system consists of two coupled channel waveguides and a slab-waveguide patch on a silica-encapsulated silicon platform (Fig. 12a). Each of the channel waveguide supports a fundamental guided mode, and the slab-waveguide patch is coupled to second channel waveguide to control the radiation loss. The time-dependent non-Hermitian Hamiltonian that governs the mode evolution $d\psi/dt = iH(t)\psi$ is taken to be traceless by a gauge transformation, that is,

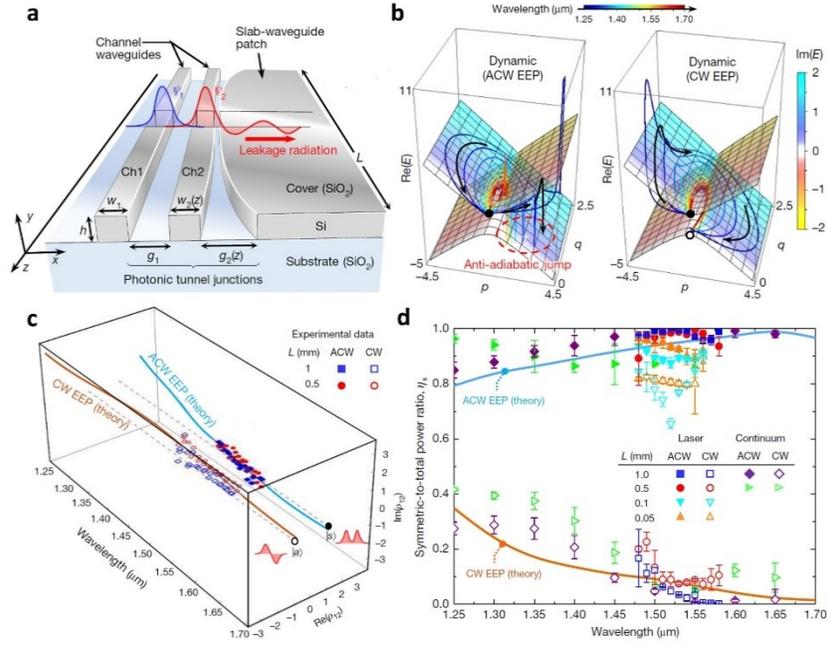

**Fig. 12 a** Schematic of the Si waveguides system. **b** Numerical results of the dynamical loop around EP in both directions. ACW loop experiences an anti-adiabatic transition while CW loop is overall adiabatic. **c** Measured complex amplitude ratio (*dots*) compared with the theory (*curves*). The output mode of ACW loop is almost the same as the input, while the output of CW loop is an asymmetric mode. **d** Measured power ratio of symmetric mode with different device lengths and different light sources (laser and supercontinuum source). Figures from Ref. [186].

$$H(t) = \begin{bmatrix} p(t) + iq(t) & 1 \\ 1 & -p(t) - iq(t) \end{bmatrix}. \tag{35}$$

Since the temporal behavior is now transferred to the propagation behavior along the waveguide system, the reduced parameters $p(t)$ and $q(t)$ denote the real and imaginary wavevectors, representing the difference in propagation constant and loss, respectively. The unitary off-diagonal terms correspond to the constant coupling rate. The parameters are controlled by the spatial profile of waveguides to form a loop that smoothly encloses the EP at $p = 0, q = 1$. The dynamics of encircling the EP is numerically computed in both anticlockwise (ACW) and CW orientations, over a broad wavelength range from 1.25 μm to 1.70 μm. Shown in Fig. 12b, the anticlockwise loops for different wavelength are all characterized by an anti-adiabatic jump and return to the initial state, while the clockwise loops result in an exchange of states despite of the excursion from the ideally adiabatic case.

In the experiment, standard nanolithography technique is used to fabricate the devices for anticlockwise and clockwise operations. The initial state is always chosen to be a symmetric state $|s\rangle = |\psi_1\rangle + |\psi_2\rangle$, and the final state is measured by fitting the diffraction patterns of the output light. The final state $|\psi_{final}\rangle = a_1|\psi_1\rangle + a_2|\psi_2\rangle$ is assessed by two quantities: the complex amplitude ratio $\rho_{12} = a_1/a_2$, and the power ratio of symmetric mode that is defined as

$$\eta_s = \left| \frac{\langle s|\psi_{final}\rangle}{\langle \psi_{final}|\psi_{final}\rangle} \right|^2 = \frac{|1 + \rho_{12}|^2}{1 + |\rho_{12}|^2}. \tag{36}$$

The measurements are repeated with devices in different lengths $L = 0.5, 1.0 \, mm$ and under various laser wavelengths ranging from 1.48 μm to 1.58 μm. The results show the complex amplitude ratio matches well with the theory (Fig. 12c). Over this 100 nm band, the final state for anticlockwise loop is mostly symmetric but for clockwise it is switched to be asymmetric, and the spectrally-averaged extinction ratio of both directions $[\eta_s]_{ACW}/[\eta_s]_{CW}$ is computed to be as large as 20.1 dB for $L = 1.0\ mm$. If the device length is reduced such that adiabaticity is always broken, the mode transfer should not happen and the experimental power ratio deviates a lot from the theoretical curve (*cyan and orange dots* in Fig. 12d, corresponding to $L = 0.1, 0.05\ mm$). Besides using the laser as excitation, a supercontinuum source is also used in the experiment. The measured power ratio under the continuum excitation exhibits the same trend of chiral mode transfer, but now in a broader spectrum that covers the entire optical telecommunication band (*purple and green dots* in Fig. 12d). The broadband asymmetric mode switching demonstrated in the Si waveguide systems shows the potential of more applications utilizing the dynamics around EPs in the optical telecommunication.

## 7. Conclusion and outlook

In this chapter, we have summarized and reviewed the initiative studies of non-Hermitian optics based on silicon photonic platforms, emphasizing throughout the fundamental physical phenomena and the associated novel applications. Transforming from quantum mechanics to classical electromagnetic waves, the broken Hermiticity of a Hamiltonian has brought forth rich opportunities for engineering photonic structures with unconventional optical behavior by judicious design of optical amplification/dissipation and coupling between those elements. The silicon photonic platforms fabricated with standard techniques, full integratability and strong scalability breed the optical structures to meet those requirements, including coupled guided wave structures with gain/loss distribution, scatterer perturbed microresonators, Brag gratings with periodic index modulation, synthetic photonic lattices, *etc*. Grounded on the rich light-matter interactions and versatile approaches of manipulating material properties in these devices and systems, numerous counterintuitive non-Hermitian properties have been revealed and demonstrated ranging from spectral singularity, chiral modes, non-orthogonal eigenstates, field localization, to unidirectional reflection, enhanced nonlinearity and exotic topology. Consequently, significant development has been spurred in applications of ultrasensitive optical sensors, unconventional lasing behavior, unidirectional and nonreciprocal light propagation, modified EIT, asymmetric mode switches, and so on.

Though vast achievement has been witnessed, we envision that much greater potential of future development of non-Hermitian silicon photonics lies in its incorporation with quantum optics domain and topological photonics.

Integrated photonic circuits fabricated on silicon chips have become an important platform that enable the implementation of quantum optical devices and networks for quantum sensing, computation and communication [187]. The introduction of non-Hermitian physics into quantum silicon photonics will not only shed interesting light on the true quantum nature of non-Hermitian notions, but also pave the way for the development of important cutting-edge quantum technology.

*PT*-symmetry has been implemented in the quantum optics domain based on an on-chip waveguide directional coupler made by the cutting-edge femtosecond direct laser writing on a fused silica wafer [188]. The directional coupler has been conventionally leveraged to generate the two photon Hong–Ou–Mandel (HOM) interference, where the two photons become indistinguishable marked by a HOM dip in the coincidence when the propagation distance equates the coupling length [189]. Here, the system renders passive *PT*-symmetry with disparate dissipation in two waveguides, where additional bending loss is introduced to the left waveguide by a slight sinusoidal modulation of the geometry. The result shows a shifted position of HOM dip to a shorter propagation distance, which is attributed to the modified quantum dynamics by the possibility of photon loss.

Up to now, the non-Hermitian physics in quantum optics remains as a nontrivial issue. There have been numerous proposals and open questions along this route. First, it has been shown that it is impossible to introduce full *PT*-symmetry to quantum optics in a similar fashion to that in classical optics, because the noise introduced by quantum gain medium will break the *PT*-symmetry [190]. However, it is possible to exploit asymmetric dissipation to realize passive *PT*-symmetry [191,192] or even leverage unitary physics of squeezing to implement non-Hermitian quantum setting in absence of dissipation [193]. Second, non-Hermitian singularity, i.e., EPs, may bring exotic influence on the quantum dynamics and fluctuation such as quantum decoherence [194,195], entanglement [196], information retrieval [197], and quantum noise behavior [81,87,88,198]. Third, it would be interesting to adapt the non-Hermitian setting to nonlinear properties of silicon/silica material, such as Kerr effect [199] and optomechanical interaction [200,201], which may bring about interesting phenomena in photon dynamics and light-matter interaction within the quantum limit [202]. With the development in silicon photonic technique especially the capability of manipulating nonclassical light such as single photon, squeezed states and entanglement, the in-depth exploration into these questions and experimental implementation will attract increasing attention and bring revenue to the fundamental study of non-Hermitian physics and quantum optics, as well as the engineering design of integrated quantum devices and networks.

Topological photonics [203,204] is an emerging branch of photonic research that revolutionizes our understanding on the transport and control of light. The essential ideas of topological photonics are adopted from the studies of electronic topological insulators [205], the latter of which is a new phase of matter with insulating bulk and conductive, robust edge states. The earliest demonstrations verifying the feasibility of topological photonics were performed in the platforms of photonic waveguide lattices [206] and resonator arrays [207]. It is not hard to anticipate that the rich physics in non-Hermitian systems will bring more intriguing stories in topological photonics, and these years have witnessed a rapid burgeoning of the research work on incorporating these two fields. $PT$-symmetric systems, as demonstrated before, can support their own topological interface modes [208], or selectively enhance the interface mode while suppressing the extended modes [209]. In a more recent experiment based on silicon photonic platform, the reappearance of interface mode via $PT$-symmetric phase transition has also been observed in a lattice of silicon waveguides [210]. A generalized theory includes these states in a bigger framework of non-Hermitian topological states [211] and predict new phase as well. In non-Hermitian cases, conventional bulk-boundary correspondence may be broken, and the topological charges associated with EPs are half-integers [212–214], and new kinds of topological transition could be realized without closing the gap or breaking usual symmetries.

On the other hand, the combination of topological photonics and active media provides innovative ideas in the design of optical components. Topological photonic lasers have been achieved in a greats variety of systems [215–221], which have advantages including robustness to scattering, high efficiency and single mode operation. Follow-up theoretical proposals predict more novel effects and applications, such as robust extended modes [222], phase transition to multimode operation [223] and frequency comb generation [224]. Another recent experiment on the reconfigurable topological light steering demonstrated the ability to control the flow of light at will [225], which could facilitate further design of optical circuitry with arbitrary light routing. Since silicon/silica materials offer convenient hybridization with gain/loss, we expect more complex optical network to be built based on silicon photonic platforms in the future.